# Mr. DLib: Recommendations-as-a-Service (RaaS) for Academia

Joeran Beel
Trinity College Dublin
ADAPT Centre, Ireland
j@beel.org

Akiko Aizawa
National Institute of Informatics
Tokyo, Japan
aizawa@nii.ac.jp

Corinna Breitinger
University of Konstanz
Germany
corinna.breitinger@uni.kn

Bela Gipp
University of Konstanz
Germany
bela.gipp@uni.kn

## ABSTRACT

Only few digital libraries and reference managers offer recommender systems, although such systems could assist users facing information overload. In this paper, we introduce Mr. DLib's recommendations-as-a-service, which allows third parties to easily integrate a recommender system into their products. We explain the recommender approaches implemented in Mr. DLib (content-based filtering among others), and present details on 57 million recommendations, which Mr. DLib delivered to its partner GESIS Sowiport. Finally, we outline our plans for future development, including integration into JabRef, establishing a living lab, and providing personalized recommendations.

## 1 INTRODUCTION

Recommender systems in academia help researchers and scientists overcome information overload. However, only a few operators of academic services, such as digital libraries and reference managers, offer recommender systems to their users. The lack of recommender systems is possibly due to the high costs of implementing and maintaining such systems. Furthermore, operators may lack the expertise to design recommender systems. In this paper, we introduce Mr. DLib's recommendations-as-a-service (RaaS), which allows operators of academic services to easily integrate recommendations in their system. For the operator, the development and maintenance effort is minimal and no expertise in designing recommender systems is required.

Only a few other companies and organizations offer recommender-systems as-a-service for academia. BibTip [15] and bX [10] are commercial RaaS provider that offer co-occurrence-based recommendations. This approach is a generic recommendation approach applicable to a variety of items (documents, movies etc.), and suitable for rather large systems with many users [3]. Additionally, coverage is rather low, i.e. the approach recommends only a fraction of the documents in a library's catalogue. CORE [13, 14] and Babel [17] offer RaaS through an API, JavaScript client, and browser plug-in. Both services recommend open-access documents (CORE indexed approx. 68 million documents, Babel approx. 40 million), and Babel states that they are welcoming researchers to evaluate novel algorithms in Babel.

While these services have their strengths and weaknesses, only Mr. DLib offers all the following features: open source; not-for-profit; supportive of research, i.e. information about the RaaS, its architecture, and effectiveness is published [5, 7, 8, 11, 16]; integration into reference managers *and* digital libraries (of any size); recommendations for open-access articles *and* the option for partners to manage private document collections; capable of recommending all documents in a corpus (100% coverage).

## 2 MR. DLIB'S RECOMMENDER SYSTEM

Mr. DLib was originally developed as a Machine-readable Digital Library at the University of California, Berkeley and introduced at JCDL 2011 [2]. Since September 2016, Mr. DLib provides recommendations-as-a-service. The concept is illustrated in Figure 1. A user browses a web site of Mr. DLib's partner, e.g. a digital library. (1) When the user looks at a specific article's detail page, the web site requests a list of related articles from Mr. DLib's RESTful Web Service as a HTTP GET request:

GET /v1/documents/{document_id}/related_documents/ [1]

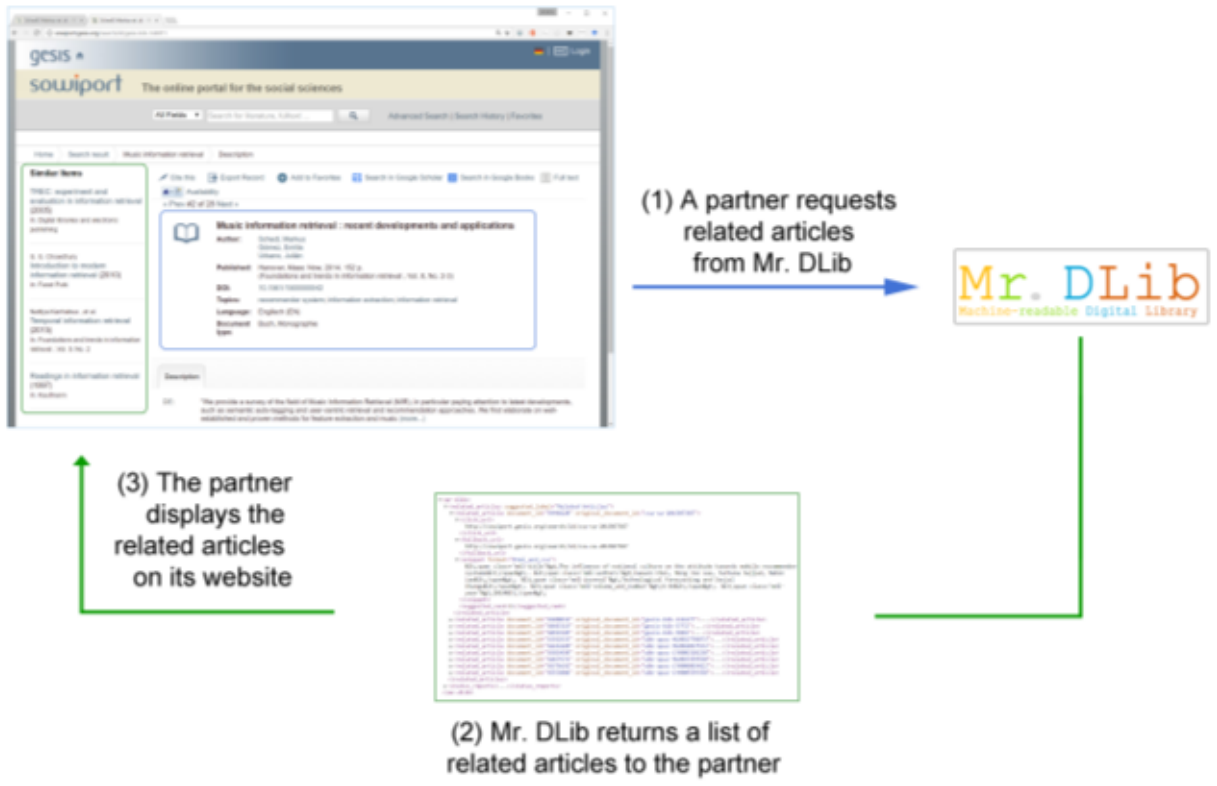

**Figure 1: Illustration of the recommendation process**

(2) Upon receiving the request, Mr. DLib computes a list of related articles, and returns the list as XML. (3) The partner website converts the XML to HTML and displays the recommendations on its web page (or mobile app, or desktop application).[2] Before this process begins, Mr. DLib indexes the metadata of the partner's documents (title, authors, abstract, venue, keywords). Mr. DLib uses Apache Lucene/Solr's *More-Like-This* function to calculate document relatedness. We are also experimenting with alternative recommendation approaches, such as stereotype and most-popular recommendations [5]. Additionally, we are experimenting with key-phrase extraction and a bibliometric re-ranking based on readership data from Mendeley [7, 8, 16].

[1] Example: https://api.mr-dlib.org/v1/documents/gesis-bib-136994/related_documents/

[2] Example: http://sowiport.gesis.org/search/id/gesis-bib-136994

The production system api.mr-dlib.org and development system api-dev.mr-dlib.org each run on a dedicated server (i7-6700k, 32GB RAM, SSD hard drives). The development system is also used for resource intensive tasks, including document indexing, key-phrase extraction, and the calculation of bibliometrics. The uptime of the servers is constantly monitored[3], and the average response time to deliver recommendations is 682ms. We further run a beta system api-beta.mr-dlib.org on a virtual machine (4 cores, 14 GB RAM).

## 3 MR. DLIB'S DIGITAL LIBRARY PARTNER

Mr. DLib's first partner is the digital library Sowiport[4], which is Germany's largest social science repository, operated by the GESIS institute [12]. Mr. DLib has indexed around 10 million documents from Sowiport. While GESIS agreed to let their documents be recommended to all partners of Mr. DLib, GESIS chose to recommend only its own content on Sowiport.

Between September 18th 2016 and February 11th 2017, Mr. DLib delivered 57,435,086 recommendations to Sowiport. Users clicked on 77,468 recommendations. This equates to an overall click-through rate (CTR) of 0.13%. This CTR is rather low, and, as shown in Figure 2, there was a notable variance (e.g. 0.19% in September and 0.10% in December). The variance may be caused by different algorithms used. In addition, recommendations are delivered when web spiders, such as the Google Bot, crawl the Sowiport website. In contrast, clicks are logged with JavaScript, which is usually not executed by web spiders.

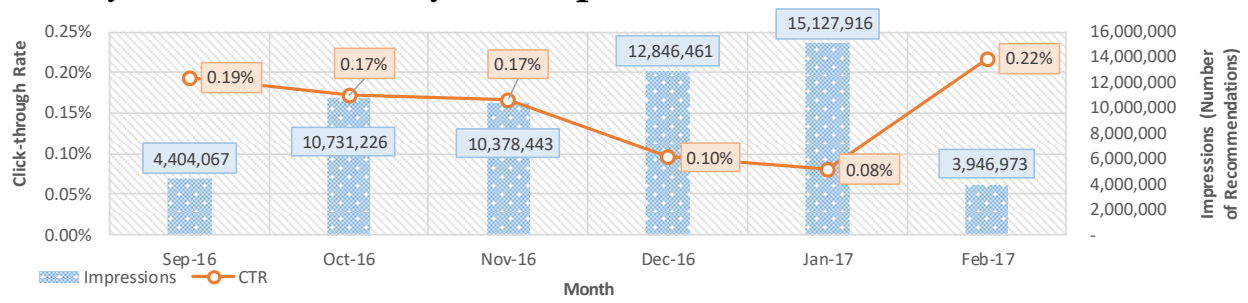

**Figure 2: Recommendations and CTR by Month**

## 4 LICENSE AND POLICY

Mr. DLib advocates an 'open culture' and publishes its code as open source on GitHub[5]. Project details are described in a public WIKI (Confluence)[6], and issues managed in a public ticket tracker (JIRA)[7]. Data from our research is published on Harvard's Dataverse[8] (if we can ensure privacy and copyrights of our partners). We invite other researchers to evaluate their recommendation algorithms with Mr. DLib and our partners[9].

## 5 OUTLOOK

In the future, we plan to add more partners (e.g. JabRef [11] and Docear [1, 4, 6]), import more documents (e.g. CORE [14]), improve the recommendation quality (e.g. cross-language recommendations [18]), and make them more personalized. We will also implement a JavaScript client that allows an easier integration of Mr. DLib's recommender system into partner websites, and monitors the number of delivered and clicked recommendations more reliably. In the long term, we will extend the scope of Mr. DLib to recommend not only related articles, but also other items relevant in academia, such as calls for papers and research grants, as well as recommending researchers as potential collaborators. In addition, we plan to establish an interactive evaluation task [9] that allows other researchers to evaluate new recommendation approaches in Mr. DLib's recommender system.

## 5 ACKNOWLEDGEMENTS

This work was supported by a fellowship within the FITweltweit programme of the German Academic Exchange Service (DAAD) and a scholarship of the Carl Zeiss Foundation. This publication has also emanated from research conducted with the financial support of Science Foundation Ireland (SFI) under Grant Number 13/RC/2106. We are further grateful for the support provided by Zeljko Carevic, Philipp Mayr, Siddharth Dinesh, Sophie Siebert, Stefan Feyer, Sara Mahmoud, Gabor Neusch, and Felix Beierle.

[3] http://monitoring.mr-dlib.org/
[4] http://sowiport.gesis.org/
[5] http://source-code.mr-dlib.org/
[6] http://wiki.mr-dlib.org/
[7] http://tickets.mr-dlib.org/
[8] http://data.mr-dlib.org/
[9] http://mr-dlib.org/research/